\documentclass[prd,superscriptaddress,twocolumn, 
preprintnumbers,showpacs]{revtex4-1} 
\usepackage{amssymb,amsmath,graphicx,epsfig} 
\usepackage{color}
\usepackage{float}
\def\beq{\begin{equation}} 
\def\eeq{\end{equation}} 
\def\bea{\begin{eqnarray}} 
\def\eea{\end{eqnarray}} 
\def\nn{\nonumber}

\def\bd{B_d} 
\def\bs{B_s} 
 
\def\bsbar{{B}_s}

\def\bstautau{ b \to s \tau^+ \tau^-} 
\def \kstar{{{K}^*}} 
\def\Bstautau{\bs \to \tau^+ \tau^-} 
\def\BKtautau{\bd \to K \tau^+ \tau^-} 
\def\BKstartautau{\bd \to \kstar \tau^+ \tau^-} 
\def\BXstautau{\bd \to X_s \tau^+ \tau^-}

\def\c{\centering}

\begin{document} 
\title{How large can the branching ratio of $B_s \to \tau^+ \tau^-$ be ?}
\author{Amol Dighe} 
\email{amol@theory.tifr.res.in} 
\affiliation{Tata Institute of Fundamental Research, 
Homi Bhabha Road, Colaba, Mumbai 400005, India} 
\author{Diptimoy Ghosh} 
\email{diptimoyghosh@theory.tifr.res.in} 
\affiliation{Tata Institute of Fundamental Research, 
Homi Bhabha Road, Colaba, Mumbai 400005, India} 
\begin{abstract} 
Motivated by the large like-sign dimuon charge asymmetry observed 
recently, whose explanation would require an enhanced decay rate of 
$B_s \to \tau^+ \tau^-$, we explore how large a branching ratio of this 
decay mode is allowed by the present constraints. We use bounds 
from the lifetimes of $B_d$ and $B_s$, constraints from the branching ratios
of related $b \to s \tau^+ \tau^-$ modes, as well as measurements of
the mass difference, width difference and CP-violating phase in the 
$B_s$-$\bar{B}_s$ system. Using an effective field theory approach, 
we show that a branching ratio as high as 15\% may be allowed while 
being consistent with the above constraints. 
The measurement of this decay, therefore, may be within the reach of current 
experiments, and can point toward a specific class of new physics models.
We also explore the possible enhancement of this decay in models with
leptoquarks and $Z'$, and find that in the latter case the branching ratio 
may be as much as 5\%, which can alleviate the dimuon anomaly to
some extent.
\end{abstract}
\preprint{TIFR/TH/12-28}
\maketitle
\section{Introduction}

In 2010, the D\O\ Collaboration reported an anomalously large
CP-violating like-sign dimuon charge asymmetry in the $B$ system
\cite{D0dimuonold-1,D0dimuonold-2}, which was strengthened by the 
updated measurement \cite{D0dimuonnew}.
The weighted average with the older CDF results \cite{CDF} gives
\bea
A_{\rm SL}^b = -(74.1 \pm 19.3) \times 10^{-4}\; , 
\label{AbSL}
\eea
which is a $3.8\sigma$ deviation from the Standard Model (SM)
prediction $A_{\rm sl}^{b \, \, {\rm SM}} = -(2.3 \pm 0.4) 
\times 10^{-4}$ \cite{Lenz}.

The measured dimuon charge asymmetry is a linear combination of the
semileptonic asymmetries $a^d_{\rm SL}$ and $a^s_{\rm SL}$ in the $B_d$
and $B_s$ sectors, respectively. Therefore, the new physics (NP) may
contribute through either of these two sectors. 
However, since
\beq
a^q_{\rm SL} = \frac{\Delta\Gamma_q}{\Delta m_q} \tan \phi_q^{\rm SL} \; 
 \quad (q = d,s) \; , 
\eeq
where $\Delta\Gamma_q$ is the width difference in the 
$B_q$-$\bar{B}_q$ system and $\phi_q^{\rm SL}$ is the CP-violating phase
in the semileptonic decays, an enhancement in $a^q_{\rm SL}$ is  
necessarily 
accompanied by an enhancement in $\tan\phi_q^{\rm SL}$ 
and/or in $\Delta\Gamma_q$.
In the $B_d$ sector, since $\tan\phi_d^{\rm SL} \approx 0.075$ in the SM 
already, a further
large enhancement would amount to fine-tuning. Also, an enhancement 
in $\Delta\Gamma_d$ would imply the NP contribution of a few percent
to the branching ratios of decay modes common to $B_d$ and $\bar{B}_d$, 
which is ruled out by the measurements of such modes \cite{HFAG}.
In the $B_s$ sector on the other hand, $\phi_s^{\rm SL} \approx 0.004$ in the 
SM \cite{Lenz}, and the branching ratios of some of the decay modes, notably
of $\Bstautau$, have not yet been strongly constrained.
Therefore NP that contributes to $\Bstautau$ would be 
a prime candidate to account for the dilepton anomaly \cite{LQs}.
Indeed, it has been shown that the only effective four-Fermi operators 
that can account for such an enhanced $A^b_{\rm SL}$ are
$(\bar{b} \Gamma s)(\bar{\tau} \Gamma \tau)$ and
$(\bar{b} \Gamma s)(\bar{c} \Gamma c)$ \cite{Bauer:2010dga}.
We, therefore, investigate how large the branching ratio 
${\cal B}(\Bstautau)$ can be.

Since the desired NP is expected to contribute to $B_s$ 
and $B_d$ decays to different extents, it may lead to a difference in the 
lifetimes of $B_d$ and $B_s$, which is otherwise expected to be 
$\lesssim 1\%$ in the SM \cite{Lenz}. 
The recent LHCb measurements \cite{LHCb:2011aa} yield 
$\tau_{B_s}/\tau_{B_d} = 1.002 \pm 0.014 \pm 0.012$.
This indicates that at the $2\sigma$ level, the branching ratio 
${\cal B}(B_s \to \tau^+ \tau^-)$ up to $3.5\%$ is still allowed, 
even if there is no NP contribution to $B_d$ decays. If NP contributes 
to $B_d$ decays, this bound will be further relaxed. 
Note that this would be a large enhancement: the value of this 
branching ratio in the SM is $\approx 7 \times 10^{-7}$ 
\cite{Grossman:1996qj}.

Measurements from other modes of the form $b \to s\tau^+\tau^-$, like
$\BXstautau, \BKtautau$ and $\BKstartautau$, can restrict the NP 
contribution to $\Bstautau$ since they involve the same 
effective four-Fermi operator.
However the experimental information on such decay modes is very poor. 
The only direct bound available at present is the 90\% BaBar 
limit of \cite{BaBarlimit}
\beq
{\cal B}(B^+ \to K^+ \tau^+ \tau^-)|_{q^2 > 14.23 \;\rm GeV^2} 
< 0.33 \% \, .
\eeq
It was suggested in \cite{Grossman:1996qj} that a dedicated experimental 
analysis of the LEP data using the missing energy spectrum as in 
\cite{Buskulic:1994gj} could be used to 
constrain $\mathcal{B}(\BXstautau)$ and $\mathcal{B}(\Bstautau)$ to 5\%; 
however such an analysis has not been performed to our knowledge. 
The bound on $\mathcal{B}(\BXstautau)$ obtained in \cite{BH} also weakens 
considerably when the theoretical uncertainties in 
$\mathcal{B}(B \to K \ell \nu + \textnormal{anything})$ are taken 
into account.
The charm counting in $B_d$ decays yields 
$\mathcal{B}(B_d \to \textnormal{no charm}) ~\leq 14\%$ 
at $2\sigma$ \cite{BH}. This would give 
$\mathcal{B}(\BXstautau) \lesssim 12.5\%$, by subtracting the SM contribution 
of 1.5\% to charmless $B$ decays \cite{Lenz:1997aa}. 
The compatibility between these bounds and a large 
${\cal B}(\Bstautau)$ needs to be investigated in a model-independent
framework as well as in the context of specific models, which we shall do 
in this paper.

The models with scalar leptoquarks \cite{Davidson:1993qk,Grossman:1996qj,LQs} 
and with flavor-dependent $Z'$ couplings \cite{Z',Li:2012xc,Kim:2012rp} 
have been proposed in the literature in order to enhance the 
$b \to s\tau^+\tau^-$ decay rates, to account for the dimuon anomaly as well 
as the measurements of width difference in the $B_s$-$\bar{B}_s$ system and 
the CP-violating phase $B_s \to J/\psi\phi$. 
We explore the two models 
above to check whether they are still able to account for all the data, at the 
same time allowing for an enhanced dimuon asymmetry.
%
{\section{Model-independent analysis: Hamiltonian for 
$B_s$-$\bar{B}_s$ mixing}}
The NP that contributes to the decay modes  $b \to s\tau^+\tau^-$ 
directly influences $B_s$-$\bar{B}_s$ mixing, 
by contributing to the dispersive ($\rm M_{12}^s$) as well as absorptive 
($\Gamma_{12}^s$) part of the effective Hamiltonian for this mixing.
These contributions may result in an enhanced value of the lifetime
difference $\Delta\Gamma_s$ and the CP-violating phase 
$\phi_s^{\rm SL} = {\rm Arg}[\rm -M_{12}^s/\Gamma_{12}^s]$ and, hence, will be 
restricted by the recent measurements of $B_s \to J/\psi \phi$ at the LHCb \cite{LHCbnew}.
In this section, we shall explore what the data have to say about
the NP contributions to $\rm M_{12}^s$ and $\Gamma_{12}^s$.

\begin{figure}[h!]
\begin{tabular}{c}
\includegraphics[width=0.8\columnwidth,height=0.6\columnwidth]{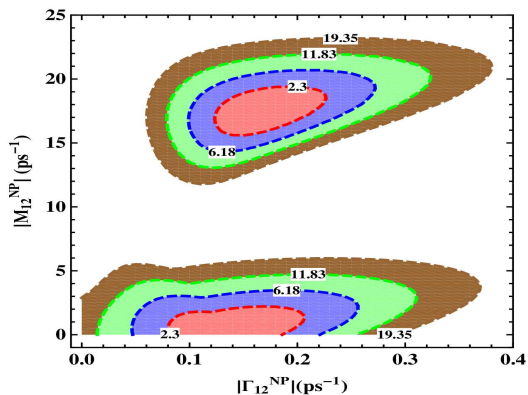}
\end{tabular}
\caption{The goodness-of-fit contours in the $|\rm M_{12}^{\rm NP}|$-$|\Gamma_{12}^{\rm NP}|$ plane, 
where the phases of $\rm M_{12}^{\rm NP}$ and $\Gamma_{12}^{\rm NP}$ are 
varied over. The red, blue, green and brown regions have $\chi^2$ values less than 
2.3, 6.18, 11.83, and 19.35, respectively, corresponding to regions allowed at 
$1 \sigma$, $2 \sigma$, $3 \sigma$, and $4 \sigma$, respectively. The same convention is 
followed in all figures.
\label{m12g12}}
\end{figure}

We follow the recent analysis \cite{Dighe:2011du}, with the
inclusion of newly available data. We parameterize the NP by
\bea
\rm M_{12}^s = \rm M_{12}^{\rm SM} + M_{12}^{\rm NP}\,, \, \,
\Gamma_{12}^s = \Gamma_{12}^{\rm SM} + \Gamma_{12}^{\rm NP} \,,
\eea  
where we take the central values of $\rm M_{12}^{\rm SM}$ and 
$\Gamma_{12}^{\rm SM}$
to be $\rm M_{12}^{\rm SM} = 8.65 \, e^{(\pi-0.04)i}$ ps$^{-1}$ and 
$\Gamma_{12}^{\rm SM} = 0.0435 \, e^{-0.04 i}$ ps$^{-1}$.
We investigate constraints on the two complex quantities 
(four parameters) $\rm M_{12}^{\rm NP}$ and  $\Gamma_{12}^{\rm NP}$ from 
the measured values of $\Delta m_s, \Delta\Gamma_s, A^b_{\rm SL}$ 
and the CP-violating phase in the $\bs \to J/\psi \phi$ decay 
($\phi^{J/\psi \phi}_s$). The results of the $\chi^2$ fit projected
in the $(|\rm M_{12}^{\rm NP}|$-$|\Gamma_{12}^{\rm NP}|)$ plane are shown
in Fig.~\ref{m12g12}. We assume all the measurements to be 
independent for simplicity, add the theoretical and experimental 
errors in quadrature, and vary over the phases of the two 
quantities $\rm M_{12}^{\rm NP}$ and $\Gamma_{12}^{\rm NP}$.

The origin in this figure is the SM, which has $\chi^2 \simeq 15$. 
This dramatically quantifies the failure of the SM to accommodate 
all the current  data.
There are two regions in the parameter space that are consistent 
with the current data, one of which is also consistent with 
$\rm M_{12}^{\rm NP}=0$, i.e. there is no need for NP to contribute
to the dispersive part of the Hamiltonian. On the other hand, 
consistency with the data even to $3\sigma$ (i.e. $\chi^2<11.83$)
seems to require a nonzero NP contribution $\Gamma_{12}^{\rm NP}$. 

The $2\sigma$ preferred range of $|\Gamma_{12}^{\rm NP}|$ is 
$(0.05,0.25)$ ps$^{-1}$. Even the lower end of this range yields 
${\cal B}(\Bstautau) \approx 15\%$ if $\Gamma_{12}^{\rm NP}$ is coming 
entirely from $\Bstautau$. 
Thus, the model-independent constraints from $B_s$-$\bar{B}_s$ mixing 
on ${\cal B}(\Bstautau)$ are rather weak and easily allow values as large as
$\sim 15\%$.
\section{Model Independent Analysis:
Hamiltonian for $b \to s\tau^+\tau^-$ decay}
\label{eff:th}
Within the SM, the effective Hamiltonian for the quark-level
transition $\bstautau$ is
\bea
\label{HSM}
&&{\cal H}_{\rm eff}^{\rm SM} = 
-\frac{4 G_{_F}}{\sqrt{2}} \, V_{ts}^* V_{tb} \, \times \\
&&\Bigl\{ \sum_{i=1}^{6} {C}_i (\mu) {\cal O}_i (\mu)
+ C_7 \,\frac{e}{16 \pi^2}\,m_b (\bar{s}
\sigma_{\mu\nu} P_R b) \, F^{\mu \nu} \nn \\
&& +\, C_9 \,\frac{\alpha_{_{em}}}{4 \pi}\, (\bar{s}
\gamma^\mu P_L b) \, \bar{\tau} \gamma_\mu \tau + C_{10}
\,\frac{\alpha_{_{em}}}{4 \pi}\, (\bar{s} \gamma^\mu P_L b) \, 
\bar{\tau} \gamma_\mu \gamma_5 \tau  \, \Bigr\}, \nn
\eea
where $P_{L,R} = (1 \mp \gamma_5)/2$. The operators ${\cal O}_i$
($i=1,..6$) correspond to the $P_i$ of Ref.~\cite{Bobeth:1999mk}, 
and $m_b = m_b(\mu)$ is the running $b$-quark mass in the 
$\overline{\rm MS}$ scheme. We use the SM Wilson coefficients as 
given in Ref.~\cite{DescotesGenon:2011yn}.

We now parameterize NP through the addition of new operators with
distinct Lorentz structures to the effective Hamiltonian for 
$\bstautau$. We consider only scalar-pseudoscalar and vector-axial vector 
operators and exclude tensor ones. 
The new effective Hamiltonian is 
\beq
{\cal H}_{\rm eff}(\bstautau) = {\cal
H}_{\rm eff}^{\rm SM} + {\cal H}_{\rm eff}^{\rm VA} + 
{\cal H}_{\rm eff}^{\rm SP}~,
\label{NP:effHam}
\eeq
where the new operators are
\bea
{\cal H}_{\rm eff}^{\rm VA} & =& 
- \frac{g_{_{\rm NP}}^2}{\Lambda^2} 
[C_V \, (\bar{s} \gamma^\mu P_L b)\, 
+  C^{\prime}_V\,(\bar{s} \gamma^\mu P_R b)]\,(\bar{\tau} \gamma_\mu \tau)\nn \\
& &  - \frac{g_{_{\rm NP}}^2}{\Lambda^2}  [C_A \, 
(\bar{s} \gamma^\mu P_L b)\,
+ C^{\prime}_A\,(\bar{s} \gamma^\mu P_R b)]\,(\bar{\tau} \gamma_\mu 
\gamma_5 \tau)~, \nn \\
{\cal H}_{\rm eff}^{\rm SP}  &=& 
- \frac{g_{_{\rm NP}}^2}{\Lambda^2} [C_S \, (\bar{s} P_R b)\,
+ C^{\prime}_S \, (\bar{s} P_L b) ]\,(\bar{\tau}\tau)
 \nn\\
&&
- \frac{g_{_{ \rm NP}}^2}{\Lambda^2} [C_P \, (\bar{s} P_R b)\,
+ C^{\prime}_P\,(\bar{s} P_L b)]\,(\bar{\tau}\gamma_5 \tau) \; .
\eea
In the above expressions, 
$g_{_{\rm NP}}$ and $\Lambda$ are the NP coupling constant and 
mass scale, respectively. The $C_i, C^{\prime}_i$ ($i=V, A, S, P$) are the unknown 
NP Wilson coefficients. Note that $C_V$ and $C^{\prime}_V$ do not contribute 
to the decay $\Bstautau$, but can contribute to the other $\bstautau$ modes. 
The indirect bounds on these NP operators 
from $b\to s\gamma$ and $b\to s\ell^+\ell^-$ ($\ell=e,\mu$) were 
considered in \cite{BH} and found to be weaker than the direct bounds 
obtained from the ratio $\tau_{B_s}/\tau_{B_d}$.

In terms of all the Wilson coefficients, the branching ratio for 
this decay is given by
\begin{eqnarray}
\label{bmumu-BR}
&& {\cal B}({\bsbar} \to \tau^+ \, \tau^-)  = 
\frac{G^2_F \alpha_{em}^2
m^5_{B_s} f_{B_s}^2 \tau_{B_s}}{64 \pi^3}
\sqrt{1 - \frac{4 m_\tau^2}
{m_{B_s}^2}}\times \nn\\
&&  
\Bigg\{
\Bigg(1 - \frac{4m_\tau^2}{m_{B_s}^2} \Bigg) \Bigg|
\xi \; \frac{C_S - C'_S}{m_b + m_s}\Bigg|^2
+ \Bigg|\xi \; \frac{C_P - C'_P}{m_b + m_s} + \nn \\ 
&&\hspace*{1.5cm}\frac{2 m_\tau}{m^2_{B_s}} 
[ (V_{tb}^{}V_{ts}^{\ast}) 
C_{10} + \xi \; (C_A - C'_A)]
\Bigg|^2 \Bigg\} \; ,
\end{eqnarray}
where $\xi \equiv (g_{\rm NP}^2/\Lambda^2) (\sqrt{2}/4 G_F)
(4\pi/\alpha_{\rm em})$.
We now examine the decays $\BKtautau$,  $\BKstartautau$ and $\BXstautau$
in the presence of NP operators. The theoretical expressions are taken 
from \cite{Alok1,Alok23}, with the replacement of $m_\mu$ with $m_\tau$. 
We first consider one NP coupling at a time. We fix the value of the 
coupling by requiring that ${\cal B} (\bsbar \to \tau^+ \tau^-) \sim 3.5\% $. 

Note that when a single NP operator is present that enhances 
${\cal B}(\Bstautau)$ to percent levels, since the SM contribution
is negligible, the branching ratio would depend only on the 
magnitude of the NP coupling and not on its phase. 
The relevant branching ratios obtained by using the values for the 
NP couplings that yield ${\cal B}(\Bstautau)=3.5\%$ are shown in 
Table~\ref{table}.
%

\begin{table}[h]
\begin{tabular}{||c|c|c|c||}
\hline
\c Model&
\c $\BKtautau$ & \c $\BKstartautau$& $\BXstautau$ \\ \hline \hline
\c SM & $0.96\times10^{-7}$ & $1.15 \times 10^{-7}$ 
& $3.6 \times 10^{-7}$ \\ \hline
\c $C_A$ & \c 0.35\;\% & \c 0.19\;\% &  0.80\;\% \\ \hline
\c $C^{\prime}_A$ & \c 0.35\;\% & \c 0.19\;\% & 0.80\;\% \\ \hline
\c $C_S$ & \c 0.04\;\% & \c 0.01\;\% & 0.05\;\% \\ \hline
\c $C^{\prime}_S$ & \c 0.04\;\% & \c 0.01\;\% & 0.05\;\% \\ \hline
\c $C_P$ & \c 0.12\;\% & \c 0.03\;\% & 0.15\;\% \\ \hline
\c $C^{\prime}_P$ & \c 0.12\;\% & \c 0.03\;\% & 0.15\;\% \\ \hline
\end{tabular}
\caption{Branching ratios of $\BKtautau$, $\BKstartautau$ and $\BXstautau$, 
when ${\cal B}(\Bstautau)= 3.5\%$ and only one type of NP coupling is present.
A kinematic cut $q^2 \geq 14.23$ GeV$^2$ has been imposed 
in our theoretical calculations for the exclusive modes. 
We have set $g_{_{\rm NP}} = g_{_{\rm EW}} = 0.65$ and 
$\Lambda= 1$ TeV. The theoretical errors on the branching ratios
may be taken to be $\sim 25\%$.}
\label{table}
\end{table}

From Table~\ref{table} we see that if there is only a single NP 
operator, the branching ratio of all these decay modes is less
than 0.8\%, and if the NP operator is scalar or pseudoscalar, it
is even less than 0.2\%. Therefore, the effect on the lifetime
ratio $\tau_{B_s}/\tau_{B_d}$ will be small. In any case, the increase 
in $B_d$ decay rate tends to increase this ratio and, hence, bring it closer to its
central value of 1.002.
It is also clear that the current 90\% upper bound on the BR of 
$\BKtautau$ is consistent with ${\cal B}(\Bstautau)$ as long as the 
single NP operator is not of the axial vector kind. 

Now we shall consider the implications of the presence of more than
one NP operator. This is of course a more realistic scenario, since
in any specific NP model, more than one effective operator is generally 
produced. For example, if the NP is leptoquark 
\cite{Davidson:1993qk,Grossman:1996qj,LQs}, in general we expect both (S 
$\pm$ P) and (V $\pm$ A) operators to be generated. And if the 
NP is a flavor-dependent $Z'$ \cite{Z'}, one obtains all the (V 
$\pm$ A) operators. 

The decay $\BKtautau$ depends on the combinations
$(C_A + C'_A), (C_S + C'_S)$, and $(C_P + C'_P)$ \cite{Alok1,Alok23}, 
while the $\Bstautau$ decay depends on the combinations
$(C_A - C'_A), (C_S - C'_S)$, and $(C_P - C'_P)$.
Therefore, it is always possible to enhance ${\cal B}(\Bstautau)$
without affecting $\BKtautau$ significantly.
On the other hand, $\BKstartautau$ depends on the same 
combination of NP operators as $\Bstautau$, except through the
$F_1$ term (see Eqs.~(A.7) and (A.12) of \cite{Alok1}) 
that depends on $(C_A + C'_A)$, but
always adds positively to the decay rate.
Therefore, the increased branching ratio of $\Bstautau$ is expected 
to also increase the branching ratio of $\BKstartautau$.
The branching ratio of $\BXstautau$ will also naturally increase.
Therefore, the latter two decay modes will be directly useful to
constrain ${\cal B}(\Bstautau)$ even when multiple NP operators
are present.

Since NP of the kind $\bstautau$ contributes to both $B_s$ and $B_d$ 
decays (through $\Bstautau$ and $\BXstautau$, respectively), the lifetime 
ratio $\tau_{B_s}/\tau_{B_d}$ can be consistent with the observation even 
if $\mathcal{B}(\Bstautau)$ and $\mathcal{B}(\BXstautau)$ are large, since 
their contributions to the 
decay widths tend to cancel each other. The bound on $\mathcal{B}(\BKtautau)$ also offers 
constraints, however, these are only marginal, due to the arguments given above. 
Note that the consistency with the bounds on $\Bstautau$, $\BXstautau$, and 
$\BKtautau$ does not need fine-tuning. For example, for $|C_A - C^{\prime}_A|$ = 3.8
(which leads to $\mathcal{B}(\Bstautau)$=15\%), any set of values of 
$|C_A + C^{\prime}_A|$ and $|C_V + C^{\prime}_V|$ in the 
range [0,1.4] are allowed by the upper bound on $\mathcal{B}(\BKtautau$). 
The value of $|C_V|$ and $|C^{\prime}_V|$ may now be chosen to be $\sim 3$ in 
order to make $\mathcal{B}(\BXstautau) \sim 12\%$, so that the $\tau_{B_s}/\tau_{B_d}$ 
constraint is satisfied. Note that further constraints on $\mathcal{B}(\BXstautau)$
would decrease the upper limit on $\mathcal{B}(\Bstautau)$ from $\sim$ 15\%.
%
\section{Leptoquark}
%
Leptoquarks (LQ) are particles whose quantum numbers are such that 
they couple to both the quarks and the leptons of the SM. Vector 
leptoquarks are predicted in many NP models like models of grand 
unification based on SU(5) \cite{Georgi:1974sy} and SO(10) 
\cite{Georgi:1974my,Fritzsch:1974nn} while scalar leptoquarks 
can arise in models of supersymmetry with R-parity violation 
\cite{Farrar:1978xj,Weinberg:1981wj,Sakai:1981pk,Aulakh:1982yn}
and in extended technicolor models \cite{Farhi:1980xs} where 
leptoquark states appear as bound states of technifermions. 
With the pre-LHCb data, a third-generation scalar leptoquark was 
shown to provide an explanation of the dimuon anomaly \cite{Dighe:2011du}. 
In this section we revisit the leptoquark explanation of the dimuon 
asymmetry in light of the recent LHCb data. We consider an SU(2) singlet 
scalar leptoquark ${\mathcal S}$ with the Lagrangian,
\beq
{\mathcal L}_{\rm LQ} \supset \lambda_{ij} \overline{(d^c)}_j P_R \ell_i 
{\mathcal S} + \rm h.c. \,,
\eeq
where $(d^c)_j$ are the charge conjugate of the down-type quark fields, 
$\ell_i$ are the charged lepton fields and $i,j$ are the generation indices.
The field ${\mathcal S}$ contributes to both $\rm M_{12}^{\rm NP}$ and 
$\Gamma_{12}^{\rm NP}$ at one loop \cite{BH} and the contribution is 
proportional to the square of the effective coupling, 
$\rm h_{_{LQ}}= \lambda^{*}_{32} \lambda_{33}$.

A 95\% C.L. lower bound of 210 GeV on the mass of a third-generation 
scalar leptoquark decaying to a $b \, \tau$ final state was reported in 
\cite{Abazov:2008jp}. 
Here, we show our results for the leptoquark mass $\rm M_{LQ} =250$ GeV.
It is worth mentioning that the above bound depends on specific assumptions 
and can be evaded if those assumptions are changed, however
our conclusions do not change even if lower masses are considered.

\begin{figure}[t!]
\begin{tabular}{c}
\includegraphics[width=0.8\columnwidth,height=0.6\columnwidth]{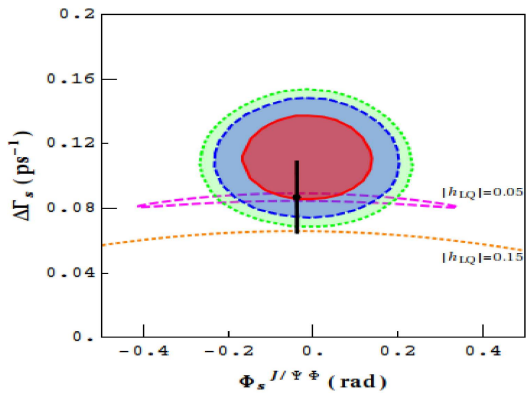}
\end{tabular}
\caption{The predictions of $(\phi_s^{J/\psi\phi},\Delta \Gamma_s)$ 
within the scalar leptoquark model
($\rm M_{\rm LQ} = 250$ GeV), overlaid on the experimental 
constraints through $B_s \to J/\psi \phi$. 
\label{lq1}}
\end{figure}

In Fig.~\ref{lq1} we show the prediction of the leptoquark model 
for two values of $|\rm h_{\rm LQ}|$ in the 
$(\phi_s^{J/\psi\phi},\Delta \Gamma_s)$ plane, superposed on the 
recent results from the LHCb \cite{LHCbnew}. The phase of
$\rm h_{\rm LQ}$ has been varied over. It is observed that 
a significant enhancement
of $\Delta\Gamma_s$ above the SM prediction is not possible in this
model.
%

A $\chi^2$ fit in the $({\rm Arg}[\rm h_{\rm LQ}],|\rm h_{\rm LQ}|$) 
parameter space, using the constraints on $\Delta m_s$ \cite{HFAG} 
in addition to those on $\Delta\Gamma_s$ and $\phi^{J/\psi \phi}_s$ 
mentioned above, yields the maximum value of $|\rm h_{\rm LQ}|$ 
to be only around 0.05 (for $\rm M_{LQ}$ = 250 GeV at 95\% C.L.). 
This gives {$\mathcal B$}($\Bstautau$) $\lesssim 0.3\%$.
A leptoquark model, thus, cannot raise to ${\cal B}(\Bstautau)$
to the level of a percent. 
It is, therefore, also not enough to explain the dimuon anomaly. 
In fact, we have checked that if $A^b_{\rm SL}$ is included 
in the fit, then the $\chi^2 \geq \chi^2_{\rm SM} \approx 15$ even 
for very low values of leptoquark mass. This means that the LQ model cannot 
do better than the SM as far as the explanation of the dimuon
anomaly is concerned.

%
\section{Flavor changing $Z^{\prime}$}
%
In this section we consider a flavor-changing $Z^{\prime}$ 
\cite{Z',Li:2012xc,Kim:2012rp} gauge boson and examine whether it can 
offer an explanation of the dimuon anomaly while still being consistent 
with the recent LHCb data. We consider the Lagrangian density
\beq
{\mathcal L}_{Z^{\prime}} \supset 
( R^{sb}[\bar{s} \gamma_\mu P_L b] Z^{\prime \mu} + \rm h.c.) 
+ R^{\tau \tau} [\bar{\tau} \gamma_\mu P_L \tau] Z^{\prime \mu}.
\eeq
Note that $\Delta m_s$ gets tree-level $Z^{\prime}$ contribution 
from ${\mathcal L}_{Z^{\prime}}$ and is proportional to 
$({R^{sb}})^2$. This restricts the value of $R^{sb}$ to be very small.
On the other hand, $\Gamma_{12}^s$ is sensitive to the 
combination of couplings $\lambda =R^{sb} R^{\tau \tau}$, and we
have already seen in Fig.~\ref{m12g12} that a large value of 
$\Gamma_{12}^{\rm NP}$ is required for a solution of the dimuon 
anomaly. This means that a large value of $\lambda$ will be needed 
in the $Z^{\prime}$ model. With a small $R^{sb}$, this could imply
a very large value of $R^{\tau \tau}$, where the calculations
would become nonperturbative and, hence, unreliable. The situation 
would become worse with increasing values of $Z^{\prime}$ mass.
Therefore we have to stay away from the nonperturbative region.

Lower experimental bounds on the $Z'$ mass can be a serious problem. 
Direct searches of pair production of $\tau^+ \tau^-$ at the Tevatron 
have provided a lower bound on the $Z'$ mass to be 399 GeV at 95\% C.L. 
\cite{Acosta:2005ij}. However since this bound assumes SM-like couplings 
of $Z'$ to all quarks, 
it can be bypassed if the $Z'$ is assumed to couple very weakly 
(or not at all) to the first-generation quarks. 
This allows us to consider a very light $Z'$, with mass as low as 
$\rm M_{Z'}=7$ GeV. Note that since ${\rm M}_{Z'} \approx {\rm M}_{B_s}$,
the decay width of $Z'$ can affect the results (at large couplings,
the width may be as large as a few GeV) and has been taken into
account in our calculations.

Note that, in principle, bounds on the effective coupling in 
$Z \to \tau^+ \tau^-$ as obtained in \cite{ALEPH:2005ab} can put 
correlated constraints on $R^{\tau \tau}$ and $M_{Z'}$. 
However, the fit therein assumes a small imaginary part of the effective 
coupling, while the one-loop correction \cite{Haisch:2011up} to 
$Z \to \tau^+ \tau^-$ through a $Z'$ exchange has a large imaginary 
part for small $M_{Z'}$. The constraints in \cite{ALEPH:2005ab} 
are, therefore, not applicable in our case. A further analysis of
$Z \to \tau^+ \tau^-$, allowing for an imaginary contribution to the
effective coupling, may yield stronger constraints.

\begin{figure}[t]
\includegraphics[width=0.8\columnwidth,height=1.3\columnwidth]{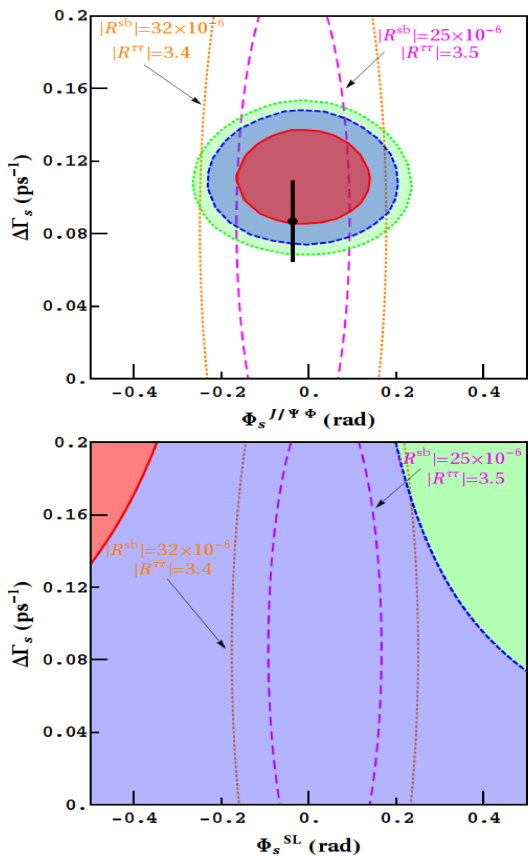}
\caption{The upper panel shows the predictions of the $Z'$ model with 
sample values of the couplings in the $(\phi_s^{J/\psi\phi} - \Delta \Gamma_s)$ 
plane overlaid with the constraints from $B_s \to J/\psi \phi$.
The lower panel shows the corresponding predictions in the 
($\phi_s^{SL} - \Delta \Gamma_s $) plane overlaid with the 
experimental constraints from $A^b_{SL}$ and $a^d_{SL}$.
We have used $\rm M_{\rm Z^{\prime}} = 7$ GeV. 
\label{zp1}}
\end{figure}

In Fig.~\ref{zp1} we show the predictions of the $Z'$ model in the 
$(\phi_s^{J/\psi\phi}, \Delta \Gamma_s)$ and 
$(\phi_s^{\rm SL}, \Delta \Gamma_s)$ planes for some sample
values of the couplings for illustration. 
(Note that $\phi_s^{J/\psi\phi} \neq \phi_s^{\rm SL}$ \cite{Lenz:2007nk,LQs}.)
The figure shows that there are values of the couplings that can be
consistent with the $J/\psi\phi$ data as well as the dimuon asymmetry data 
to within $2\sigma$.
The values of ${\cal B}(\Bstautau)$ in these allowed regions
can be as high as $\sim$ 20\%. 
However, constraints from the measurements of $\tau_{B_s}/\tau_{B_d}$, 
$\mathcal{B}(\BXstautau)$ and $\mathcal{B}(\BKtautau)$ still continue to apply. 
In the case of a specific model such as this, the constraints would be more severe 
than those in Sec.~\ref{eff:th} since the effective couplings ($C^{\prime}_V, C^{\prime}_A$) are 
now related to each other. 

\begin{figure}[t!]
\centering
\includegraphics[width=0.8\columnwidth,height=0.6\columnwidth]{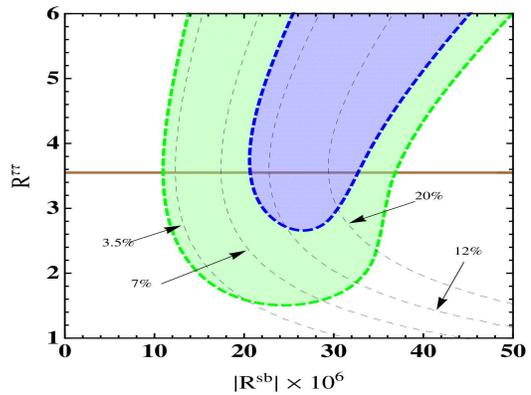}
\caption{Goodness-of-fit contours in the 
$(|R^{sb}| - R^{\tau \tau})$ plane for $\rm M_{Z^{\prime}}=7$ GeV. 
The horizontal line corresponds to $R_{\tau\tau} = \sqrt{4\pi}$.
Contours corresponding to ${\cal B}(\Bstautau)$ values of 3.5\%, 7\%, 12\%, 
and 20\% are also shown.
\label{zp2}}
\end{figure}

Figure~\ref{zp2} shows the result of a $\chi^2$ fit in the 
$(|R^{sb}|, R^{\tau \tau})$ plane using all the experimental data: 
$\Delta m_s$, $\Delta\Gamma_s$, $A^b_{\rm SL}$,seem consistent with all data
to within $2\sigma$, the results are reliable only when the 
couplings are perturbative, i.e. 
$\alpha_{_{R^{\tau \tau}}}\equiv (R^{\tau \tau})^2/4\pi < 1$.  
We indicate this value of $R_{\tau\tau}$ by a horizontal line
in the figure, and consider only the region below this line as
the valid one. (The region below the line but very close to it 
may still have significant higher-order corrections.)

%

Even by demanding perturbative couplings in addition to $\chi^2<6.18$
(i.e. $2\sigma$ level), the branching ratio 
${\cal B}(\Bstautau)$ can be as high as $\sim 20\%$. Thus, 
the $\Delta m_s, J/\psi\phi$, and $A^b_{\rm SL}$ data by itself
cannot put strong constraints on the value of this branching ratio.
The strongest constraints here come from $\tau_{B_s}/\tau_{B_d}$ and 
$\mathcal{B}(\BKtautau)$. Taking into account these constraints, the value of 
$\mathcal{B}(\Bstautau)$ can be as high as 5\%. This can happen, for 
example, with $R^{sb} = 15 \times 10^{-6}$, $R^{\tau \tau} = 3.0$, which makes   
$\mathcal{B}(\BXstautau)$ $\approx$ 1.5\% and $\mathcal{B}(\BKtautau)$ $\approx$ 0.35\%, 
the former allowing a relaxation of the constraint from $\tau_{B_s}/\tau_{B_d}$.

Note that this value of ${\cal B}(\Bstautau)$ can alleviate the dimuon 
anomaly (from $\chi^2=15.1$ for the SM, to $\chi^2=9.5$); however, it is 
not enough to explain it. Some contribution from NP in the $B_d - \bar{B_d}$ 
mixing may also be needed, as was recently conjectured in \cite{Lenz:2012az}.

\section{Concluding remarks}
%

We have investigated the possible enhancement of ${\cal B}(\Bstautau)$
that may help explain the observed anomalous like-sign dimuon asymmetry.
Taking into account the constraints from the lifetime ratio of $B_s$ 
and $B_d$, as well as the measurements of related $b \to s\tau^+ \tau^-$
decay modes, we find that an enhancement up to 
${\cal B}(\Bstautau) \sim 15\%$ is allowed at $2\sigma$. 
This bound may decrease 
with further constraints on $\mathcal{B}(\BXstautau)$.

Within the context of specific models, an enhancement of 
$\mathcal{B}(\Bstautau)$ to 
more than 0.3\% is not possible with the leptoquark model, 
but the model with a light flavor-changing $Z'$ that does not 
couple to light quarks can increase it to 5\%. 
This helps alleviate the $A^b_{\rm SL}$ anomaly to some extent,
but cannot account for it entirely, and contribution from NP
in the $B_d$ sector may be needed.

Our results with leptoquarks are similar to those in \cite{BH},
however the updated LHCb data used by us has made the constraints
on leptoquark parameters even stronger.
On the other hand, the light $Z'$ employed by us allows much larger
values of the branching ratio than that predicted therein.
Accounting for the dimuon anomaly to within $1\sigma$, 
as indicated in \cite{Z'}, is not possible even with a
light $Z'$, because of the new stronger LHCb constraints and the
requirement of perturbative couplings.

If $\mathcal{B}(\Bstautau)$ is at the percent level, it will soon be 
within the reach 
of experiments, and could play an important role in our understanding of
physics beyond the SM. The direct measurement of this branching ratio should, 
hence, be a high-priority.
%

\acknowledgments

We thank A. K. Alok, C. Bobeth, R. Godbole, M. Grunewald, U. Haisch, 
A. Kundu, A. Lenz, D. London, S. Nandi, and S. K. Patra for 
valuable discussions and comments.
%

\end{document}